\documentclass[
 amsmath,
 amssymb,
reprint,
floatfix
]{revtex4-2}

\usepackage{graphicx}
\usepackage{dcolumn}
\usepackage{bm}
\usepackage{graphicx}
\usepackage{sidecap}
\usepackage{caption, subcaption}
\usepackage{units}
\DeclareUnicodeCharacter{2212}{-}
\usepackage[T1]{fontenc}

\usepackage{textgreek}
\usepackage{color,soul} 
\usepackage{comment}
\usepackage{float} 
\usepackage{physics}
\usepackage{soul}

\usepackage{mathtools, nccmath}
\usepackage{dashrule}

\newcommand{\beginsupplement}{%
        \setcounter{table}{0}
        \renewcommand{\thetable}{S\arabic{table}}%
        \setcounter{figure}{0}
        \renewcommand{\thefigure}{S\arabic{figure}}%
        \setcounter{section}{0}
        \renewcommand{\thesection}{S\arabic{section}}%
     }

\begin{document}


\title{Pressure tuning of HgCdTe epitaxial layers - the role of the highly disordered  buffer layer}

\author{D.~Yavorskiy}
\email[]{Dmitriy.Yavorskiy@fuw.edu.pl}
\affiliation{Institute of High Pressure Physics, Polish Academy of Sciences, ul. Sokołowska 29/37, 01-142 Warsaw, Poland}
\affiliation{CENTERA, CEZAMAT, Warsaw University of Technology, ul. Poleczki 19, 02-822 Warsaw, Poland}
\affiliation{Institute of Physics, Polish Academy of Sciences, Aleja Lotników 32/46, PL-02-668 Warszawa, Poland}
 
\author{Y.~Ivonyak}
\affiliation{Institute of High Pressure Physics, Polish Academy of Sciences, ul. Sokołowska 29/37, 01-142 Warsaw, Poland}
 
\author{D.~But}
\affiliation{Institute of High Pressure Physics, Polish Academy of Sciences, ul. Sokołowska 29/37, 01-142 Warsaw, Poland}
\affiliation{CENTERA, CEZAMAT, Warsaw University of Technology, ul. Poleczki 19, 02-822 Warsaw, Poland}

\author{K. ~Karpierz}
\affiliation{Faculty of Physics, University of Warsaw, ul. Pasteura 5, 02-093 Warsaw, Poland.}
 
\author{A.~Krajewska}
\affiliation{Institute of High Pressure Physics, Polish Academy of Sciences, ul. Sokołowska 29/37, 01-142 Warsaw, Poland}
\affiliation{CENTERA, CEZAMAT, Warsaw University of Technology, ul. Poleczki 19, 02-822 Warsaw, Poland}
 
\author{M.~Haras}
\affiliation{Warsaw University of Technology, Centre for Advanced Materials and Technologies CEZAMAT, ul. Poleczki 19, 02-822 Warsaw, Poland}
\affiliation{Gdańsk University of Technology, Faculty of Electronics, Telecommunications and Informatics, Department of Functional Materials Engineering, ul. G. Narutowicza 11/12, 80-233 Gdańsk, Poland.}

\author{P.~Sai}
\affiliation{Institute of High Pressure Physics, Polish Academy of Sciences, ul. Sokołowska 29/37, 01-142 Warsaw, Poland}
\affiliation{CENTERA, CEZAMAT, Warsaw University of Technology, ul. Poleczki 19, 02-822 Warsaw, Poland}
 
\author{M.~Dub}
\affiliation{Institute of High Pressure Physics, Polish Academy of Sciences, ul. Sokołowska 29/37, 01-142 Warsaw, Poland}
\affiliation{CENTERA, CEZAMAT, Warsaw University of Technology, ul. Poleczki 19, 02-822 Warsaw, Poland}
 
\author{A.~Kazakov}
\affiliation{International Research Centre MagTop, Institute of Physics, Polish Academy of Sciences, Aleja Lotników 32/46, 02-668 Warsaw, Poland}

\author{G.~Cywi\'nski}
\affiliation{Institute of High Pressure Physics, Polish Academy of Sciences, ul. Sokołowska 29/37, 01-142 Warsaw, Poland}
\affiliation{CENTERA, CEZAMAT, Warsaw University of Technology, ul. Poleczki 19, 02-822 Warsaw, Poland}

\author{W.~ Knap}
\affiliation{Institute of High Pressure Physics, Polish Academy of Sciences, ul. Sokołowska 29/37, 01-142 Warsaw, Poland}
\affiliation{CENTERA, CEZAMAT, Warsaw University of Technology, ul. Poleczki 19, 02-822 Warsaw, Poland}
 
\author{J.~ {\L}usakowski}
\affiliation{Faculty of Physics, University of Warsaw, ul. Pasteura 5, 02-093 Warsaw, Poland.}

\date{\today}

\begin{abstract}

Hg$_{1−x}$Cd$_x$Te alloys are unique because by increasing the Cd content $x$, one modifies the band structure from inverted to normal, which fundamentally modifies the dispersion of bulk and surface or edge (in the case of quantum wells) energy states. Using alloys with $x$ close to the concentration $x_c$ at which the band inversion transition is observed and with additional application of hydrostatic pressure ($p$), one creates a favorable playground for studying the evolution of Dirac matter and its topological properties. In this work, cryogenic magnetospectroscopy in  quantizing magnetic fields ($B$) at the far-infrared is used to study inter-Landau-level transitions in high-quality Hg$_{1−x}$Cd$_x$Te MBE-grown epitaxial layers with $x \approx x_c$  as a function of $p$ up to 4.2 kbar. Special attention is paid to elucidate the role of the substrate and buffer layers, which usually modify the pressure coefficients of epitaxial layers. For this purpose, comparative measurements were carried out on as-grown epilayers with a GaAs substrate and on free-standing layers obtained by etching off the substrate. Spectra registered as a function of $B$ (at given $p$) were analyzed with the help of the Kane model modified to include magnetic field. The pressure coefficient as well as the difference between conduction and valence band deformation potentials of the free-standing layer were determined at 2 K. Surprisingly, the deformation potentials and pressure coefficients of the epitaxial layer and those of the free-standing layer differed by no more than 10\% in the pressure range up to 4.2 kbar. This finding questions the common belief of a dominant influence of the substrate on the pressure coefficients of epitaxial layers. We attribute the smallness of this difference to the presence of a highly disordered CdTe buffer separating the substrate from the epitaxial layer, which relaxes the transmission of strain from the substrate to the layer. Our results contribute to a better understanding of pressure experiments carried out on epitaxial layers on a substrate.

\end{abstract}


\maketitle

\section{\label{sec:level1}Introduction}

Research on Hg$_{1-x}$Cd$_x$Te (HgCdTe, for short) crystals dates back to the late 1950s, when their sensitivity to mid-infrared and far-infrared radiation was discovered \cite{WDLawson_1959}. The research on these crystals, carried out in the second half of the XX$^{\mbox{th}}$ century, led to  development of detectors that are currently among the most sensitive and reliable detectors in these ranges of the electromagnetic spectrum. As high-sensitivity and low-noise devices, they have been installed, for example, in the J. Webb telescope in the form of a multi-pixel matrix \cite{JDGarnett_2004}. A review of   results of this broad and long-lasting research carried out on HgCdTe detectors can be found in \cite{RDornhaus_1976} and \cite{ARogalski_2005}.

Despite such a long history of research, HgCdTe remains a widely and actively studied material. From the point of view of fundamental physics, the main reason for the recent important interest in HgCdTe-based 3D and 2D structures comes from the fact that HgCdTe technology allows for the creation of a Dirac matter in these materials, i.e., a solid-state system in which the dependence of energy versus momentum of carriers is linear and which is accompanied by surface or edge dissipation-less conductivity (a topological transport). It has also been shown that electronic states in bulk HgCdTe can exhibit properties of pseudo-relativistic three-dimensional particles \cite{MOrlita_2014, FTeppe_2016}.

The evolution of these specific properties can be analyzed by tuning the energy band gap with the alloy composition, $x$. Indeed, by increasing $x$, one moves (at the $\Gamma$ point of the Brillouin zone) from an inverted-band HgTe to the direct band-gap CdTe passing a zero-gap semiconductor at $x \approx$ 17\% (if we consider liquid helium temperatures). What is more, tuning of the band gap can be achieved at a constant value of $x$ with temperature or pressure \cite{MOrlita_2014, FTeppe_2016, NBBrandt_1976, MS_2022}.

It has been recently shown that at about 10 GPa, HgCdTe undergoes a phase transition and becomes superconducting \cite{Ya-KangPeng_2021}; this discovery may start a new direction in the research on these crystals. Also, in ~\cite{MS_2022} a convincing result of the application of hydrostatic pressure to tune the band structure of HgCdTe epitaxial layers on a GaAs substrate close to the topological (inverted-to - direct band gap) transition was presented.

The epitaxial growth of 3D or 2D materials brings research inevitably to the question of the role of the substrate, which is especially important in pressure experiments. However, until now, it has not been clear how substrates modify the pressure coefficients of a HgCdTe epitaxial layer.

It is known (see Table \ref{tab:bulk_modulus}) that HgCdTe is a more compressible material than GaAs, Si, or Ge, and its bulk modulus is noticeably smaller.

\begin{table}[ht!]
\caption{Literature data on bulk modulus.}
	\begin{center}
	\begin{tabular}{|c|c|c|c|c|}
		\cline{2-4}
		\multicolumn{1}{c|}{} & \multicolumn{3}{|c|}{Bulk modulus [GPa]} \\
		\hline
		{Material} & {2 K/4.2 K} & {80 K} & {300 K}  & {Ref.} \\
		\hline
		Si		&	& & 97.8 & \cite{HJMcSkimin_1953}\\
		\hline
		Ge		&	& & 75.1 & \cite{HJMcSkimin_1953}\\
		\hline
		GaAs		& 76.87 & & 75.28 & \cite{RICottam_1973}\\
		\hline
		ZnTe		&  & &  50.8 & \cite{BHLee_1969}\\
		\hline
		CdTe  	    & 45.312 &  & 43.05 & \cite{SRajagopalan_1979}\\
		\hline
		HgTe		& 46.96  & & 42.27 & \cite{RICottam_1975}\\
		\hline
		Hg$_{1-x}$Cd$_x$Te: 0$\le$x$\le$1	& & 44-47 & & \cite{DBagot_1994}\\
		\hline
	\end{tabular}
	\label{tab:bulk_modulus}
	\end{center}
\end{table}

In this work, we address the question of the influence of substrate on the pressure coefficients of the band structure in the case of high-quality Hg$_{0.85}$Cd$_{0.15}$Te epitaxial layers MBE-grown on GaAs substrate with the bulk modulus almost two times higher than that of HgCdTe alloys. For this purpose, we carried out comparative measurements on as-grown sample and on sample from which the substrate was removed (free-standing layer). This way, in the pressure range up to 4.2 kbar, we were able to determine  the pressure coefficients of a free-standing  layer as well as the difference of the deformation potentials of the conduction and valence band of this layer to be $a_c − a_v$ = -3.06 eV. Surprisingly, the difference of pressure coefficients of the epitaxial layer on the substrate $a_c − a_v$ = -2.77 eV differs by no more than 10\% from these of the substrate-free layer which questions the common belief of a dominant influence of a hard substrate on the pressure coefficients of soft epitaxial layers. We attribute this small difference in pressure coefficients to the presence of a highly disordered CdTe buffer separating the substrate from the epitaxial layer, which play the role of a "damper", softening the transmission of strain from the substrate to the layer. Our results are of importance in analyzing pressure dependent properties of HgCdTe-based  systems.

\section{Samples and experimental set-up}

\begin{figure*}[ht!]
     \centering
     \captionsetup{justification=centering,margin=0.1cm}
     \begin{subfigure}{0.4\textwidth}
         \includegraphics[width=\textwidth]{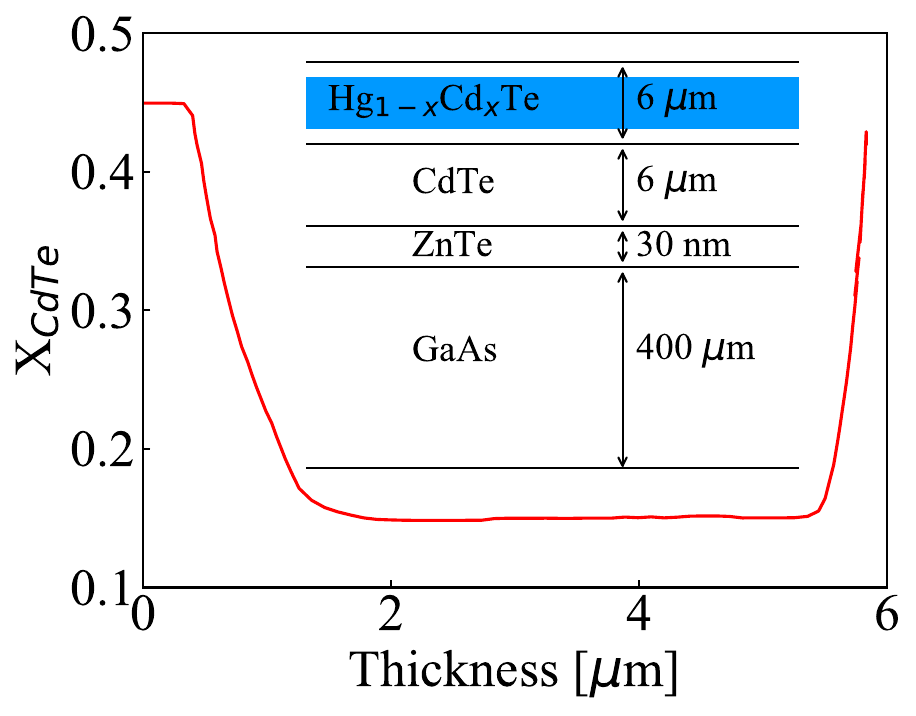}
         \caption{}
         \label{fig:cd_profile}
     \end{subfigure}
     \hfill
     \begin{subfigure}{0.45\textwidth}
         \includegraphics[width=\textwidth]{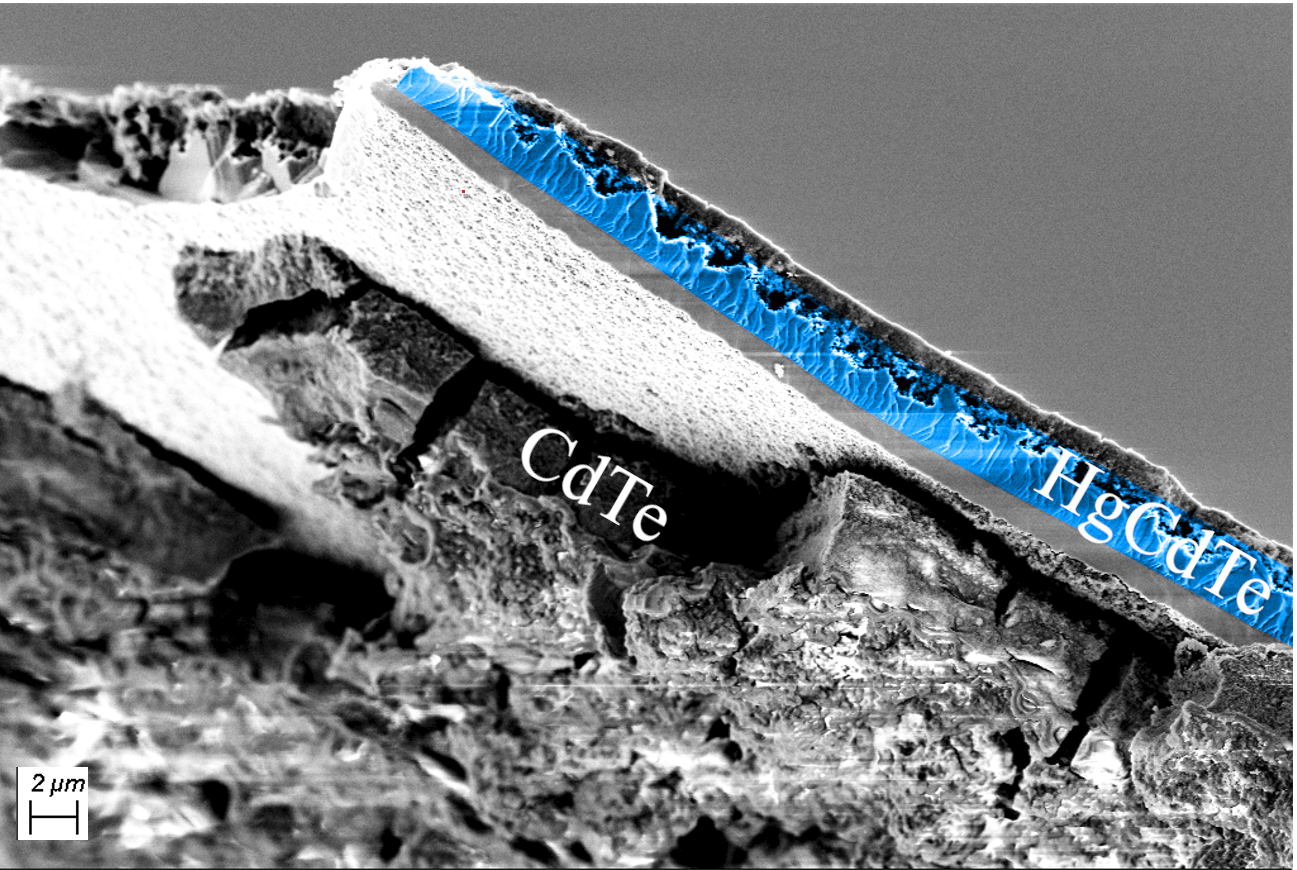}
         \caption{}
         \label{fig:pic}
     \end{subfigure}
\caption{(a) Cd content in Hg$_{0.85}$Cd$_{0.15}$Te layer as a function of the distance from the interface between the Hg$_{0.85}$Cd$_{0.15}$Te and CdTe layers. Inset: a scheme of the structure of the sample studied. (b) A cross section of HgCdTe flake (SEM); the blue area corresponds to the studied layer with $x$ = 0.15. A black, cracked layer is a CdTe buffer.}
\end{figure*}

An MBE-grown HgCdTe epitaxial layer with $x$ = 0.15 was studied in magnetospectroscopy experiments under hydrostatic pressure. The structure of the sample is shown in Fig.~\ref{fig:cd_profile}: a 400~$\mu$m-thick substrate of semi-insulating GaAs, a 30-nm-thick ZnTe layer, a 6~$\mu$m-thick CdTe buffer, and a 5.85~$\mu$m-thick layer of HgCdTe layer with the Cd composition profile shown in Fig.~\ref{fig:cd_profile}; the Hg$_{0.85}$Cd$_{0.15}$Te layer with a thickness equal to 3.65~$\mu$m is covered with a 0.5~$\mu$m-thick cap layer of HgCdTe with Cd content ranging from 15\% to 45\%.

The sample was studied in two different forms: one was as-grown, while the substrate was removed from the other sample by etching in a solution of H$_2$O$_2$ and NH$_4$OH followed by rinsing in distilled water and drying in air. In this way, a flake with a thickness of about 12~$\mu$m (6~$\mu$m of CdTe and 6~$\mu$m of HgCdTe) and a size of 1~mm~x~1~mm was obtained. Even though its thickness was only 12~$\mu$m, the flake was sufficiently stiff and could be placed in a pressure chamber. A cross-sectional scanning electron microscopy (SEM) image of a detached flake is shown in Fig.~\ref{fig:pic}.

In the pressure experiments, the samples were closed in an optical pressure cell filled with a 1:1 mixture of transformer oil and petroleum spirit as the pressure medium. To facilitate a transfer of the sample without substrate to the pressure cell, it has been put into a Teflon capsule with holes that allow the mixture to penetrate inside. Experiments were carried out with the cell cooled down to 2~K. The pressure cell used in these experiment allows to achieve up to $\sim$10 kbar at 300 K. The maximum pressure applied to the sample at room temperature was 7.2~kbar; this pressure is about two times lower than the critical pressure (of about 15~kbar, \cite{SNarita_1973}) at which the phase transition of HgCdTe from the zinc blend to the hexagonal cinnabar crystal structure takes place. While cooling to liquid helium temperatures, the pressure is typically reduced by about 40\% which results in the maximum pressure at 2~K equal to only 4.2~kbar.

Transmission of monochromatic electromagnetic radiation with a wavelength of 118.8~$\mu$m (2.52~THz) was studied as a function of the magnetic field. A molecular laser pumped with a CO$_2$ laser was the source of radiation, and the transmitted signal was registered with a carbon bolometer placed just below the sample in the pressure cell. The spectra (the signal from the bolometer registered as a function of the magnetic field) were normalized to the power of the laser beam measured with a pyroelectric detector.


\section{Results and discussion}

\begin{figure}[ht!]
     \centering
     \begin{subfigure}{0.475\textwidth}
         \includegraphics[width=\textwidth]{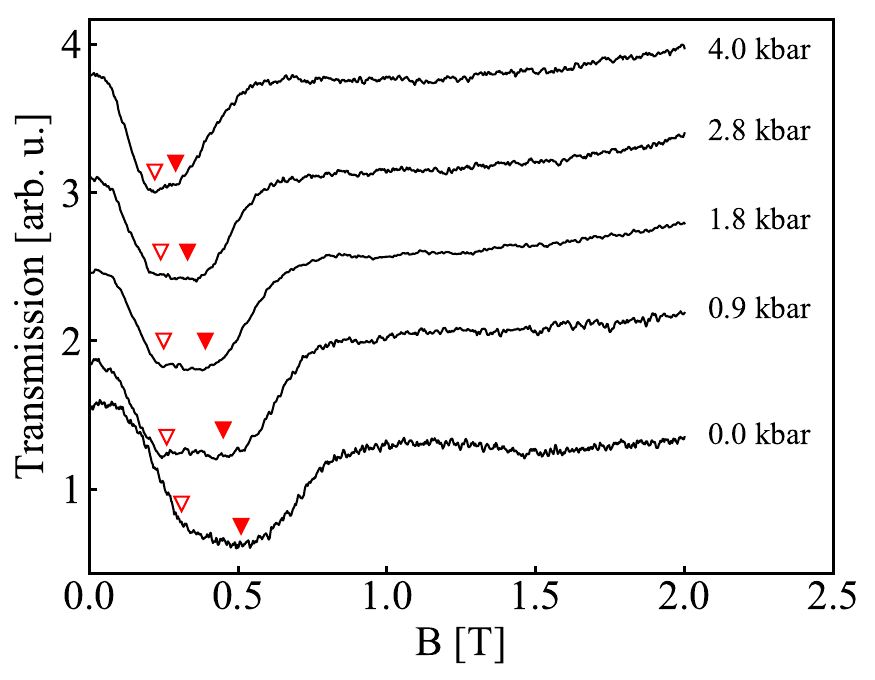 }
         \caption{}
         \label{fig:tr_118_8mum_dif_press_190402}
     \end{subfigure}
     \hfill
     \begin{subfigure}{0.475\textwidth}
         \includegraphics[width=\textwidth]{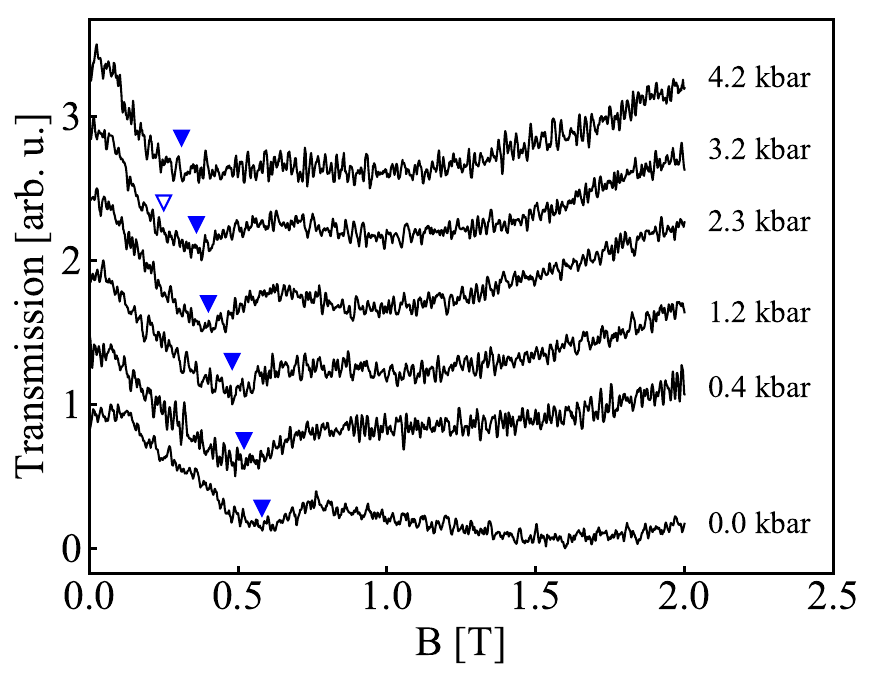}
         \caption{}
         \label{fig:tr_118_8mum_dif_press_190402_no_GaAs}
     \end{subfigure}
	\caption{Transmission spectra for 118.8~$\mu$m radiation at different pressures (indicated in the figures) for Hg$_{0.85}$Cd$_{0.15}$Te layer on GaAs substrate (a) and without the substrate (b). Red and blue triangles indicate resonances, whose positions are plotted as a function of magnetic field in Fig. \ref{fig:dysp_118_8mum_190402_yes_no_GaAs_diff_press}.}
\end{figure}

\begin{figure}[ht!]
     \centering
         \includegraphics[width=0.475\textwidth]{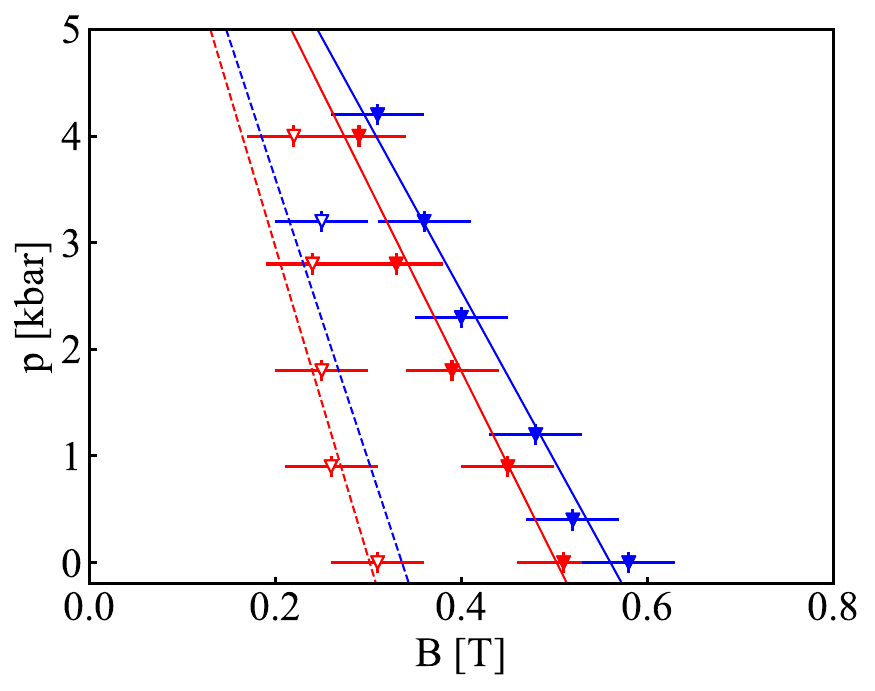}
	\caption{Positions of resonances at different hydrostatic pressures obtained from Figs. \ref{fig:tr_118_8mum_dif_press_190402}, 
 and \ref{fig:tr_118_8mum_dif_press_190402_no_GaAs} for the Hg$_{0.85}$Cd$_{0.15}$Te layer on the GaAs substrate (red triangles) and without the substrate (blue triangles). Red and blue solid lines represent linear fits of LLT$_2$ only, while dotted lines correspond to the LLT$_1$ transition, which is plotted based on fitting of LLT$_2$.}
	\label{fig:dysp_118_8mum_190402_yes_no_GaAs_diff_press}
\end{figure}

Magnetotransmission spectra are shown in Fig.~\ref{fig:tr_118_8mum_dif_press_190402} and Fig.~\ref{fig:tr_118_8mum_dif_press_190402_no_GaAs} for the epitaxial layer with and without the substrate, respectively.  The spectra were normalized to their value at zero magnetic field and shifted along the $y$ axis for clarity. Positions of resonances are indicated with red and blue triangles and  presented in Fig.~\ref{fig:dysp_118_8mum_190402_yes_no_GaAs_diff_press} on the pressure - magnetic field plot. The spectra for the epitaxial layer without GaAs substrate are much noisier than the spectra for the layer on the substrate, and the presence of the resonances at lower magnetic fields is only visible at~3.2~kbar. A worse signal-to-noise level at resonances for the sample without substrate may result from both the smaller size of the HgCdTe flake and the Teflon capsule in which the sample was sealed. In both cases, the absorption coefficient for 118.8~$\mu$m at second resonance is equal to around $6\times10^3$~cm$^{-1}$.

Optical properties of HgCdTe at energies close to the fundamental band gap are well described by the Kane model ~\cite{EvanOKane_1957} and its extensions, which include, in particular, the presence of a magnetic field (e.g., see Ref.~\cite{MHWeiler_1977}). In the present work, we base the analysis on a simplified Kane model, which is discussed in details in the Supplementary Materials to Ref.~\cite{MOrlita_2014}. According to this model, the dependence of the energy of $n^{\mbox{th}}$ Landau level at zero component of the momentum  in the direction of $B$ $p_z = 0$ on the magnetic field is given by a pseudo-relativistic formula:

\begin{equation}
\begin{split}
\begin{multlined}
E_{\xi, n, \sigma}(p_z)=\xi^2\widetilde{m}\widetilde{c}^2+(-1)^{1-\theta(\widetilde{m})}\xi\sqrt{\widetilde{m}^2\widetilde{c}^4+}\\[0.5ex]
\overline{\rule{0pt}{2.4ex} + \frac{e \hbar \widetilde{c}^2B}{2}(4n-2+\sigma)+\widetilde{c}^2p_z^2}
\end{multlined}
\end{split}
\label{f1}
\end{equation}

\noindent where $\widetilde{c}$ is the Kane fermion velocity, which is a constant equal to about $1.05\times10^6$ m$/$s, $\widetilde{m}$ is the rest mass of the particle (electron or hole), and is related to the bandgap energy $E_G$ by the relation $E_G$ = 2 $\widetilde{m}$ $\widetilde{c}^2$. The number $n$ indexes Landau levels and takes positive integer values; $\xi$ describes electron, light-hole, and heavy-hole bands if equal to +1, -1, and 0, respectively; $\sigma$ is a quantum number related to the projection of the spin on the magnetic field and can take values $\pm$ 1.

\begin{figure}[ht!]
     \centering
         \includegraphics[width=0.475\textwidth]{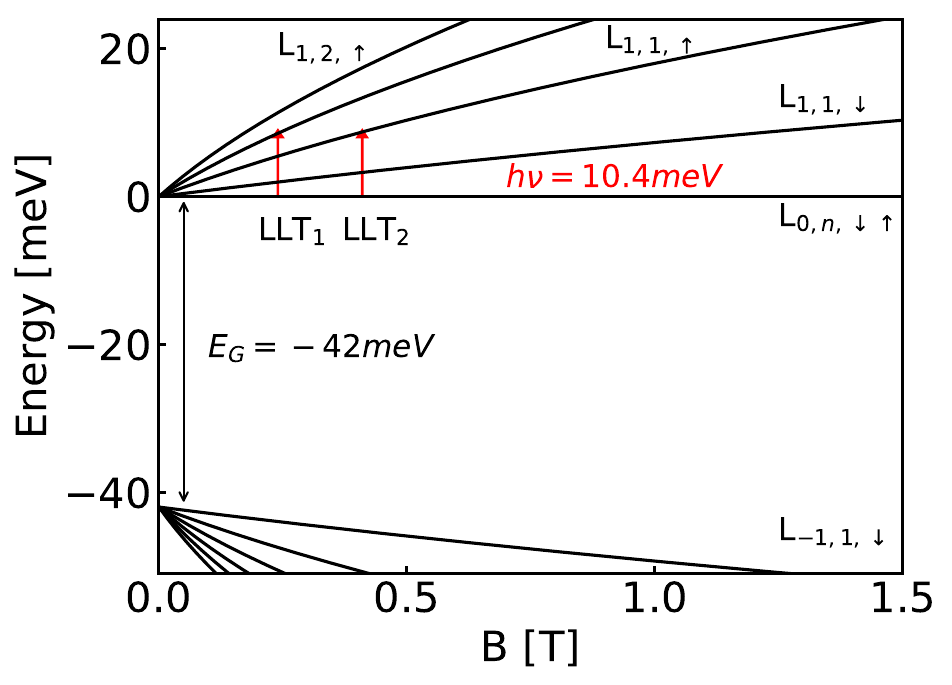}
	\caption{Landau levels chart for the Hg$_{0.85}$Cd$_{0.15}$Te for $E_G$ = -42 meV.}
	\label{fig:LL}
\end{figure}

According to this model and selection rules for the electric dipole transitions ($\Delta n$ = 0 and $\Delta \sigma$ = 0), the two resonances observed in the magnetotransmission spectra can be interpreted as transitions between Landau levels of the heavy hole band and Landau levels of the conduction band described by Eqs.~\ref{r1} and \ref{r2}. They are shown schematically with red arrows in Fig.~\ref{fig:LL} for $p$ = 0 and $E_G$ = -42~meV. These transitions were also observed in \cite{MOrlita_2014} and \cite{FTeppe_2016}.

\begin{equation}
LLT_1:\ LL_{\xi =0, n=3, \sigma=-1} \rightarrow LL_{\xi =1, n=2, \sigma=-1}
\label{r1}
\end{equation}
\begin{equation}
LLT_2:\ LL_{\xi =0, n=2, \sigma=1} \rightarrow LL_{\xi =1, n=1, \sigma=1}
\label{r2}
\end{equation}

In order to determine the pressure coefficient of the energy gap, we assumed that $E_G(p) = E_G(0) + \alpha p$. A linear dependence of $E_G$ on $p$ in the pressure range of interest is justified by the results of Ref.~\cite{SNarita_1973} and \cite{IMTsidilkovski_1985}. As it was demonstrated for the detached sample, the only clearly visible resonance was interpreted as the LLT$_2$ transition. At lower magnetic fields, the LLT$_1$ transition is hardly visible, so we decided to analyze only LLT$_2$ resonance. A linear fitting procedure applied to the data in Fig.~\ref{fig:dysp_118_8mum_190402_yes_no_GaAs_diff_press} gave the parameters $\alpha$ and $E_G$ shown in Table \ref{tab:fit}. 

\begin{table}[ht!]
        \caption{Parameters $\alpha$, $E_G$, and $a_c-a_v$ obtained from fitting.}
	\begin{center}
	\begin{tabular}{|c|c|c|c|}
		\cline{2-4}
		\multicolumn{1}{c|}{} & \multicolumn{1}{|c|}{$\alpha$ [$\nicefrac{\mbox{meV}}{\mbox{kbar}}$]} & \multicolumn{1}{c|}{$E_G$ [meV]}& \multicolumn{1}{|c|}{$a_c-a_v$ [eV]}\\
		\hline
		With substrate &	5.9 $\pm$ 1.2 &	-42 $\pm$ 1 & 2.77 $\pm$ 0.56\\
		\hline
		Without substrate &	6.5 $\pm$ 1.3 &	-48 $\pm$ 1 & 3.06 $\pm$ 0.61\\
		\hline
	\end{tabular}
	\label{tab:fit}
	\end{center}
\end{table}

For the sample with the substrate, we obtained $E_G \approx$ -~~42~meV which is close to the value calculated from an empirical formula determining the dependence of the band gap on temperature and cadmium composition: $E_G$(2~K, 15\%)~=~-33~meV \cite{JPLaurenti_1990}. Removal of the substrate caused an increase in $E_G$ by about 14\% and an increase in $\alpha$ by about 10\%. The pressure coefficients obtained in the present study are similar, and are within the range of values for HgCdTe determined by other research groups in the past, as Table~\ref{tab:alfa} shows.

\begin{figure}[ht!]
     \centering
         \includegraphics[width=0.475\textwidth]{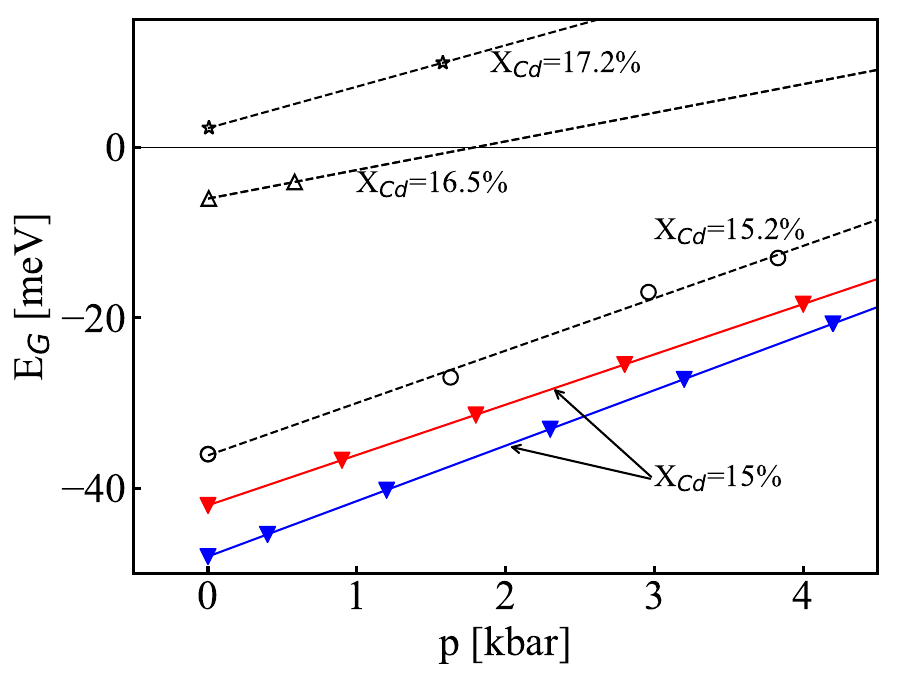}
	\caption{Plot of $E_G$ vs. $p$, red and blue triangles are for the HgCdTe epilayer on the GaAs substrate and the free-standing epilayer, respectively. Open black symbols are from   \cite{MS_2022} and concern the HgCdTe epitaxial layers with $x$ equal 15.2\%, 16.5\%, and 17.2\%.}
	\label{fig:EgvsP}
\end{figure}

In Fig.~\ref{fig:EgvsP}, the dependence of $E_G$ on $p$ is shown in comparison with data from~\cite{MS_2022}. It can be seen that the pressure coefficients for different Cd-content epilayers are similar to those obtained from the present work.

The pressure modification of the band gap of HgCdTe layers is determined by the values of the deformation potential of conduction ($a_c$) and valence bands ($a_v$). The values of $a_c$ and $a_v$ cannot be directly determined experimentally, but their difference, $a_c - a_v$, can be extracted from the hydrostatic pressure dependence of the direct band gap $E_G(P)$ or $dE_g/dp$ \cite{MS_2022}. Using the theoretical approach developed in \cite{MS_2022} we can estimate the $a_c - a_v$ as

\begin{equation}
\frac{\partial E_G}{\partial p}\bigg|_{p=0} = -\frac{1}{B_0(T)}(a_c - a_v),
\end{equation}

\noindent where $B_0(T)$ is the temperature-dependent bulk modulus which was  taken to be $B_0$ = 47 GPa \cite{MS_2022}. This leads to the difference of the deformation potentials for the free-standing HgCdTe layer, $a_c − a_v$ = -3.06 eV at 2 K (see Table \ref{tab:fit}). This is, to the best of our knowledge, the first experimental estimation  of $a_c - a_v$ for a free-standing HgCdTe layer.

In our opinion, the most interesting result of the study is a relatively small difference in the pressure coefficient $\alpha$ between the samples with and without the substrate. This seems to be in contradiction with a simple argument saying that deformation of an epitaxial layer should follow deformation of a much harder and thicker substrate - see bulk modulus of GaAs, HgCdTe and CdTe in Table~\ref{tab:bulk_modulus}. Then, removal of the substrate should essentially change the coefficient $\alpha$, but this effect was not observed in the experiment.

To explain the relatively small difference of $\alpha$ for the two $x$ = 0.15 samples studied in the pressure experiments, we refer to known data on the role of the CdTe buffer in separating the GaAs substrate from the CdTe epitaxial layers ~\cite{KKarpierz_2008}. There it was shown by means of magnetotransport and THz magnetospectroscopy that even a relatively thick layer of CdTe of the thickness of about 5~$\mu$m on a GaAs substrate is of a very low crystallographic quality because of a large lattice mismatch between GaAs and CdTe. A scanning electron microscopy picture of the CdTe buffer layer in the sample measured in the present work after removing the substrate is presented in Fig.~\ref{fig:pic}. What is more, in Ref. \cite{DWasik_2001} it was shown that application of hydrostatic pressure leads to the formation of stacking faults, dislocations, and other structural defects in the CdTe/Cd$_{1-x}$Mg$_x$Te structure in the vicinity of the II–VI/GaAs interface. The reason of creation of these defects is referents of the compressibility between a GaAs and II–VI layers.

We propose that this cracked buffer layer (see Fig. \ref{fig:pic}) is acting as a ''damper'', which does not transfer all the elastic properties of the substrate to the epitaxial layer. It serves as kind of  mechanical isolation that allows the epitaxial layer to show pressure coefficients largely independent of a thick and more rigid substrate. This phenomenon was observed in other systems in which a large lattice mismatch between a substrate and an epitaxial layer is present. For instance, in Ref.~\cite{AKaminska_2016}, a high-pressure photoluminescence was studied at varying tickness of Zn$_{1-x}$Mg$_x$O layers, and the authors showed that the pressure coefficients depended mainly on the thickness of the layer and hence on the presence of strain in the film. They observed that for relaxed layers, pressure coefficients (for different $x$) are higher than for strained layers. 

This qualitative statement should be made quantitative by carrying out calculations that would take into account the elastic parameters of GaAs substrate, CdTe buffer, and HgCdTe layer. This would, however, require theoretical modeling of the pressure properties of randomly disordered CdTe buffer grown on GaAs. This is, however, beyond the scope of the present paper. 

\begin{table}[ht!]
        \caption{Literature data on the coefficient $\alpha$ of Hg$_{1-x}$Cd$_{x}$Te. 1~-~epilayer on GaAs, 2~-~free standing epilayer.}
	\begin{center}
	\begin{tabular}{|c|c|c|c|c|}
		\hline
		$x$ [\%]&$\alpha$ [$\nicefrac{meV}{kbar}$]&Technique&$T$ [K] &Ref.\\
		\hline
		0		&	10.4$\pm$0.6	&Transport& \> 60 & \cite{JStankiewicz_1976}\\
		\hline
		12.4    &	9.45			&Transport&4.2& \cite{MMGDeCarvalho_1983}\\
		\hline
		15		&	3				&Transport&77& \cite{CFau_1984}\\
				&   7				&Transport&77& \cite{CTElliott_1972}\\
                &   6.1				&Optical&2& \cite{MS_2022}\\
		  &	5.9$^1$ & Optical & 2 &this work\\
 		  &	6.5$^2$ & Optical & 2 &this work\\
		\hline
		0$\leq x \leq$15  &	10$\pm$2&Transport &2$\leq$T$\leq$300&\cite{NBBrandt_1976}\\
		\hline   	 	
		20   	&	9.5  			&Electrical&4.2 & \cite{JunhaoChu_1994}\\
		\hline 
		0$\leq$x$\leq$30  &	12$\pm$2&Transport &\> 77& \cite{CVerie_1966}\\
		\hline
		70		&	8.7				&Optical   &300 & \cite{SJiang_1992}\\
		\hline 
		100     &   8.3             &Optical   &300 & \cite{WeiShan_1985}\\
				&	8				&Optical   &300 & \cite{JRMei_1984}\\
				&   7.9$\pm$0.2     &Optical   &80  & \cite{HMCheong_1991}\\
				&   8.4  			&Optical   &300 & \cite{JGonzalez_1995}\\
				&   6.5$\pm$0.2  	&Optical   &2   & \cite{DJDunstan_1988}\\
		\hline
	\end{tabular}
	\label{tab:alfa}
	\end{center}
\end{table}

\section{Conclusions}

In this work, we present low temperature studies on HgCdTe epilayers with the dispersion of the  band structure close to the topological phase transition. By combining THz magnetotransmission with application of hydrostatic pressure, we studied the influence of the substrate on the pressure coefficients of an MBE-grown HgCdTe layer on a semi-insulating GaAs substrate. Comparative magnetospectroscopy measurements carried out on as-grown epilayers with a GaAs substrate and on free-standing layers were analyzed with the help of a modified Kane model. 

The band-gap pressure coefficient  $\alpha = 6.5$~meV/kbar as well as the difference of deformation potentials of the conduction and valence band  $a_c − a_v$ = -3.06 eV  for a free-standing epilayer were determined at 2 K up to 4.2 kbar. Unexpectedly, the deformation potentials and pressure coefficients of the epitaxial layer on the substrate and the free-standing  epitaxial layer (after removal the substrate) were found to be very close. This questions the common belief of a the strong influence of the substrate on the pressure coefficients of epitaxial layers. We attribute this relatively small differences to the presence of a highly disordered CdTe buffer separating the substrate from the epitaxial layer, which relaxes the transmission of the substrate strain to the layer. Our results are of general importance for interpretation of pressure experiments carried out on epitaxial layers on a substrate.

\begin{acknowledgments}

The authors are thankful to N.~N.~Mikhailov, S.~A.~Dvoretsky, and V.~I.~Gavrilenko for providing the studied crystals and helpful discussions.

This research was partially supported by a Polish National Science Center UMO-2019/33/B/ST7/02858 grant, the International Research Agendas Program of the Foundation for Polish Sciences co-financed by the European Union under the European Regional Development Fund (CENTERA: No. MAB/2018/9), the European Union through ERC-ADVANCED grant TERAPLASM (No. 101053716). Views and opinions expressed are, however, those of the author(s) only and do not necessarily reflect those of the European Union or the European Research Council Executive Agency. Neither the European Union nor the granting authority can be held responsible for them. We also acknowledge the support of "Center for Terahertz Research and Applications (CENTERA2)" project (FENG.02.01-IP.05-T004/23) carried out within the "International Research Agendas" program of the Foundation for Polish Science co-financed by the European Union under European Funds for a Smart Economy Programme. D.B.B. acknowledges the funding from the European Union's Horizon 2020 research and innovation program under grant agreement No. 857470. This research was also supported by the Foundation for Polish Science through the Projects from IRA Programme co-financed by EU within SG OP and FENG (project “MagTop” no. FENG.02.01-IP.05-0028/23). Publication subsidized from the state budget within the framework of the programme of the Minister of Science (Polska) called Polish Metrology II project no. PM-II/SP/0012/2024/02, amount of subsidy 944,900.00 PLN, total value of the project 944,900.00 PLN.

\end{acknowledgments}

\bibliography{bib}
\bibliographystyle{apsrev4-2}

\beginsupplement

\end{document}